\documentclass[doublecol,figures]{epl2}

\usepackage{amssymb,amsmath}
\usepackage{epsf,colordvi,graphicx,color,amsbsy}
\usepackage[T1]{fontenc}

\title{Observation of depth-induced properties in wave turbulence on the surface of a fluid}
\author{E. Falcon\thanks{Corresponding author: \email{eric.falcon@univ-paris-diderot.fr}} \and C. Laroche} 
\shortauthor{E. Falcon \and C. Laroche}
\institute{Laboratoire Mati\`ere et Syst\`emes Complexes (MSC), Universit\'e Paris Diderot, CNRS -- UMR 7057\\ 10 rue A. Domon \& L. Duquet, 75 013 Paris, France, EU
}


\abstract{We report the observation of changes in the wave turbulence properties of gravity-capillary surface waves due to a finite depth effect. When the fluid depth is decreased, a hump is observed on the wave spectrum in the capillary regime at a scale that depends on the depth. The possible origin of this hump is discussed. In the gravity regime, the wave spectrum still shows a power law but with an exponent that strongly depends on the depth. A change in the scaling of the gravity spectrum with the mean injected power is also reported. Finally, the probability density function of the wave amplitude rescaled by its rms value is found to be independent of the fluid depth and to be well described by a Tayfun distribution.}
\begin{document}

\maketitle
 
\section{Introduction} 
Wave turbulence is ubiquitous in nature. It ranges from surface or internal waves in oceanography,  Alfv\'en waves in solar winds, Rossby waves in geophysics, elastic or spin waves in solids, waves in optics, and quantum waves in Bose condensates (for reviews see~\cite{Falcon,Newell}). Wave turbulence theory (a statistical theory describing an ensemble of weakly nonlinear interacting waves) predicts analytical solutions of the kinetic equations of weak turbulence at the equilibrium or in a stationary out-of equilibrium regime in various systems involving wave dynamics \cite{ZakharovLivre}. Surprisingly, well-controlled laboratory experiments on wave turbulence were scarce \cite{Wright96} up to last years where new observations was reported such as intermittency \cite{Falcon07b}, fluctuations of the energy flux \cite{Falcon08}, finite size effect of the system \cite{Falcon07,Denissenko07} and the full space-time power spectrum of wave amplitudes \cite{Herbert10}.  Several questions still are open, notably about the validity domain of the theory in experiments (horizontal finite size effects, role of strongly nonlinear coherent structures), and the possible existence of weak turbulence solutions for nondispersive (ie, for 2D acoustic waves) or weakly dispersive systems \cite{Newell71,Connaughton03}. Indeed, the lack of dispersion could lead to cumulative nonlinear effects leading to shock wave formation. It is thus of primary interest to know the evolution of the wave energy spectrum when the dispersion relation of a system is changed from a dispersive to a nondispersive regime.

A simple way to experimentally do this is to change the depth of a fluid on which surface waves propagate. Indeed, gravity waves are known to become nondispersive in a shallow water limit. In this limit, weak turbulence predicts power spectra of wave amplitude much less steep than in the deep regime for both gravity \cite{Zakharov99,Onorato09} and capillary \cite{Kats74} wave turbulence. The prediction for the gravity regime has been tested in few laboratory experiments using a large-scale wave flume with a sloping bottom \cite{Kaihatu07,Smith03}.  At a more applied level, in situ observations exist in oceanography when ocean waves propagate from deep water into shallow coastal areas \cite{Smith03}. Notably, when ocean surface waves approach the shore, bottom friction and depth-induced wave breaking are no more negligible, and near-resonant triad interactions could play a dominant role instead of the 4-wave interaction process \cite{Wiserevue07,Ochi05}.
 
In this letter, we study gravity-capillary wave turbulence on the surface of a fluid within a constant depth tank. When the depth is decreased from a deep to a thin fluid layer, a hump is observed on the power spectrum in the capillary regime at a scale that depends on the depth. In the gravity regime, the wave spectrum still shows a power-law but with an exponent that depends on the depth. The scaling of the power spectrum with the mean injected power is measured for different fluid depths as well as the probability distribution of the wave amplitudes. This latter is found to be independent of the fluid depth and to be well described by a Tayfun distribution.

\section{Experimental set-up}
\begin{figure}[t!]
\includegraphics[width=8.7cm]{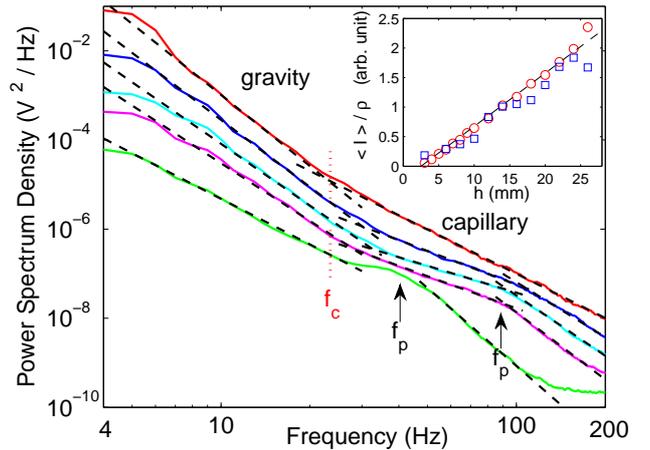} 
\caption{(color online) Power spectra of wave amplitude $S(f)$ for different fluid depths from $h=22$, 11, 8, 6 and 4 mm (solid lines from top to bottom). Dashed lines are spectrum slopes reported on Fig. 2. $f_c$ is the crossover frequency between gravity and capillary regime and $f_p$ the hump frequency (see text). Forcing bandwidth and amplitude: 1 -- 4 Hz and $\sigma_V=0.8$ (arb. unit). Curves have been vertically shifted for clarity by a factor 1, 1.5, 2.5, 6 and 12 (from bottom to top). Inset: Mean injected power rescaled by the fluid density $\langle I \rangle / \rho$ versus $h$ for water ($\square$) and mercury ($\circ$). Dashed line is a linear fit. Forcing parameters: 1 - 6 Hz and $\sigma_V= 0.8$  (arb. unit).}
\label{fig01}
\end{figure}

The experimental setup, described in \cite{Falcon07}, consists of a rectangular plastic vessel,  20 $\times$ 20 cm$^2$, filled with mercury (unless otherwise stated) up to a depth $h$. $h$ is varied from 3 mm up to 22 mm with a 0.1 mm accuracy. Mercury is used because of its low kinematic viscosity $\nu= 10^{-7}$ m$^2$/s. Surface waves are generated on the fluid by the horizontal motion of a rectangular ($L \times H$ cm$^2$) plunging plastic wave maker driven by an electromagnetic shaker. We take $L=13$ cm and $H=3.5$ cm, the wave maker being at 0.2 mm from the bottom vessel. The wave maker is driven with low-frequency random vibrations (typically from 1 to 5 Hz). The amplitude of the surface waves $\eta(t)$ at a given location is measured by a capacitive wire gauge (plunging perpendicularly to the fluid at rest) \cite{Falcon07}.  $\eta(t)$ is analogically low-pass filtered at 1 kHz and is recorded during 300 s by using an acquisition card with a 2 kHz sampling rate. The power injected $I(t)$ into the fluid by the wave maker is also measured using a force sensor and a velocity sensor both fixed on the wave maker \cite{Falcon08}. The system is driven by the rms value of the velocity fluctuations $\sigma_V$ of the wave maker \cite{Falcon07}. The mean power injected by the wave maker $\langle I \rangle$ is measured as a function of the depth $h$ for the same value of the forcing amplitude $\sigma_V$. As shown in the inset of Fig.\ \ref{fig01} (for both water or mercury as the working fluid), one finds that $\langle I \rangle \sim \rho h$ for $h>h^*=2.5$ mm. $\rho$ is the density of the fluid (the mercury density being 13.6 times the water one). $h^*$ is the depth below which the dewetting of the fluid on the vessel bottom occurs that forms islands of fluid surrounded by domains without fluid. One has also $\langle I \rangle \sim \rho h \sigma_V^2$ (not shown here). Thus, typical experiments are performed either keeping constant the forcing amplitude $\sigma_V$  when changing the fluid depth, or adjusting $\sigma_V$ to have a constant mean injected power $\langle I \rangle$ when the depth is decreased. We have performed experiments with two different fluids (mercury or water). Experimental results with mercury are shown below, but qualitative similar results have been observed with water.

\section{Dispersion relation of linear waves}
For an arbitrary fluid depth $h$, the dispersion relation of inviscid linear surface waves on a fluid reads
\begin{equation}
\omega^2=(gk+\frac{\gamma}{\rho}k^3)\tanh (kh)
\label{rdtheo}
\end{equation}
where $\omega$ is the wave pulsation, $k$ its wavenumber, $g=9.81$ m/s$^2$ the acceleration of the gravity, $\gamma=0.4$ N/m the surface tension, and $\rho=13.6 \times 10^{3}$ kg/m$^3$ the density of the mercury. Gravity waves are dominant in Eq.\ (\ref{rdtheo}) for wavelengths $\lambda \gg 2\pi l_c \simeq$ 1 cm [capillary length $l_c\equiv \sqrt{\gamma/(\rho g)}\simeq$ 1.7 mm] whereas capillary waves dominates when $\lambda \ll 2\pi l_c$. The deep-water limit corresponds to $kh\gg 1$ that is for $\lambda \ll 2\pi h$, whereas the shallow-water limit corresponds to $\lambda \gg 2\pi h$. Typically, for our range of frequencies $4 \leq f \leq 200$ Hz, Eq.\ (\ref{rdtheo}) leads to $0.1 \leq \lambda / (2\pi h) \leq 2.3$ for $h=3$ mm and $0.01 \leq \lambda / (2\pi h) \leq 0.7$ for $h=20$ mm.  One thus probes rather $\lambda \sim 2\pi h$ for lowest frequencies and the smallest depths than $\lambda \gg 2\pi h$.  One focuses below on wave turbulence (involving nonlinear interacting waves) from a deep to a thin fluid layer. 

\begin{figure}[t]
\includegraphics[width=8.5cm]{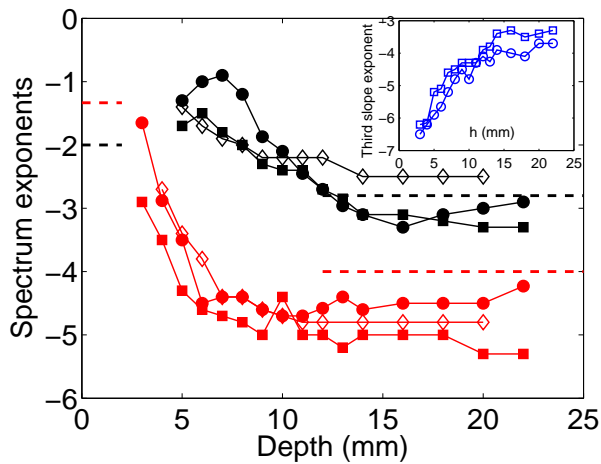}
\caption{(color online) Exponents of power spectra as a function of the depth for gravity [red (light grey) symbols] and capillary (black symbols) regimes. Forcing parameters: 1 - 4 Hz ($\square$), 1 - 6 Hz ($\circ$) both at $\sigma_V= 0.8$  (arb. unit), or at constant $\langle I \rangle=0.02$ (arb.\ unit) ($\lozenge$). Dashed lines are theoretical exponents for both regimes in the deep or shallow water limits (see text). Inset: Third exponent of the spectrum as a function of $h$ fitted between $f_p$ and the noise level frequency.}
\label{fig02}
\end{figure}

\section{Wave turbulence power spectrum}
The power spectra of wave amplitude $S(f)$ are shown in Fig.\ \ref{fig01} for different depths $h$ at a constant forcing amplitude $\sigma_V$. For large depths (e.g. $h=22$ mm), it displays similar results than those found previously in the deep regime: two power laws corresponding to the gravity and capillary wave turbulence regimes separated by a crossover frequency $f_c$ related to $l_c$ \cite{Falcon07}. When $h$ is decreased, important changes are observed in the spectrum shape (see Fig.\ \ref{fig01}). First, a hump appears on the power spectrum at a frequency denoted $f_p$ (see Fig.\ \ref{fig01}). $f_p$ is found to decrease when $h$ is decreased. Second, the power spectrum still shows a frequency-power law in the gravity regime (frequencies below $f_c$) but with a strongly decreasing exponent. Third, in the capillary regime (between $f_c$ and $f_p$), a power-law can be also fitted (except for $h=4$ mm) in a narrower inertial range due to the presence of the hump (see Fig.\ \ref{fig01}). Finally, at small enough depth, a third power-law can be fitted on the high frequency part of the spectrum (between $f_p$ and a frequency given by the intercept of the spectrum and the signal to noise level - see Fig.\ \ref{fig01}). This power-law vanishes when the deep regime is reached. Similar observations are found either when performing experiments at constant $\langle I \rangle$ forcing instead of constant $\sigma_V$ forcing, or when using an optical non-invasive method (laser vibrometer) instead of the capacitive wire gauge. Moreover, these observations do not significantly depend on the probe location on the basin surface except close to the boundaries. 

Let us now find the scaling of the spectrum amplitude $S(f)$ with $\langle I \rangle$ for a fixed depth. In deep regime ($h=20$ mm), $S(f)$ is found to scale as $\langle I \rangle^{0.8\pm0.1}$ for both gravity and capillary regimes on almost one decade in $\langle I \rangle$. This result is consistent with ones found in previous measurements \cite{Falcon07,Xia10} whereas weak turbulence predicts a scaling in $\langle I \rangle^{1/3}$ for 4-wave interactions (gravity in deep regime) and in $\langle I \rangle^{1/2}$ for 3-wave interactions (capillary regime) \cite{ZakharovLivre}. For small depths ($h=5$ mm), $S(f)$ is found to scale as $\langle I \rangle^{0.5\pm0.1}$ for gravity waves, whereas no scaling is found for the capillary regime due to the presence of the hump. The observed change of scaling with the depth thus suggests that a change of the wave-interaction process occurs for gravity waves.
 
\begin{figure}[t!]
\includegraphics[width=8.5cm]{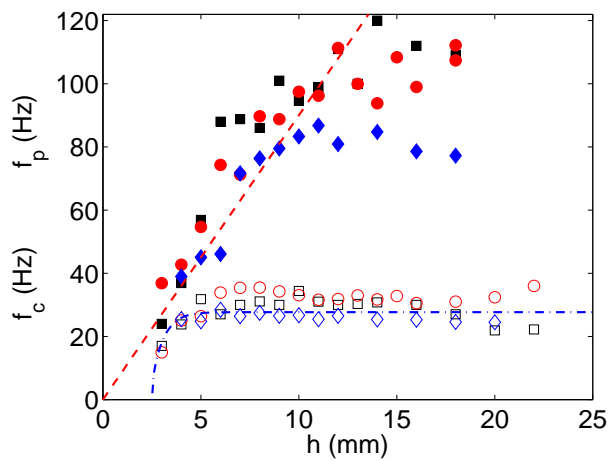} 
\caption{(color online) Hump frequency $f_{p}$ (full symbols) and crossover frequency $f_{c}$ (open symbols) as a function of the fluid depth $h$ for different forcings: 1 - 4 Hz ($\square$), 1 - 6 Hz ($\circ$) both at $\sigma_V= 0.8$  (arb. unit), or  at constant $\langle I \rangle=0.02$ (arb.\ unit) ($\lozenge$).  Dashed line: linear fit of slope 9 Hz/mm. Dot-dashed line is given by the depth dependence of the dispersion relation at a fixed wavenumber $\tilde{k}_c=1.56\sqrt{\rho g/\gamma}$ (see text).}
\label{fig03}
\end{figure}

Figure\ \ref{fig02} shows the evolution of the frequency exponents of power spectra $S(f)$ as a function of the depth for the gravity and capillary regimes for different forcing parameters. When $h$ is decreased, the absolute values of exponents continuously decrease both for the capillary and gravity regimes. The decrease of the exponents when $h$ decreases is qualitatively consistent with the expected one from the weak turbulence predictions in both depth limits. Indeed, weak turbulence predicts in both depth limits: $S_{grav}(f) \sim f^{-4}$ (deep)  \cite{Zakharov67Grav} and $\sim f^{-4/3}$ (shallow) \cite{Zakharov99,Onorato09} for the gravity regime (see dashed-red [light grey] lines in Fig.\ \ref{fig02}), and $S_{cap}(f) \sim f^{-17/6}$ (deep) \cite{Zakharov67Cap} and $f^{-2}$ (shallow) \cite{Kats74} for the capillary regime (see dashed-black lines in Fig.\ \ref{fig02}). The power-law modification of the gravity regime could be related to a change of wave interaction process with $h$ (see above). For the capillary regime, it is related to the presence of the hump (see below). Finally, as shown in the inset of Fig.\ \ref{fig02}, when $h$ is increases, the exponent of the third power-law ending the cascade is found to strongly decreases in absolute value, and then saturates to roughly the same exponent as the capillary one.

 \section{Crossover and hump frequencies}
 Figure\ \ref{fig03} shows the evolutions of the crossover frequency $f_c$ between gravity and capillary regimes and of the hump frequency $f_p$ as a function of $h$ for different forcing parameters. $f_p$ is measured on the power spectrum as the frequency corresponding to the end of the power-law capillary cascade and the beginning of the third slope (see Fig.\ \ref{fig01}). This cut-off $f_p$ is found to strongly increase with $h$ at small depths, and then seems to saturate at larger depths ($h\geq 10$ mm). Contrarily to $f_c(h)$ (see below), the evolution of $f_p(h)$ is not given by the depth dependence of the dispersion relation of Eq.\ (\ref{rdtheo}) at any fixed wavenumber. Moreover, the evolution of $f_p(h)$ is not due to the fact that less and less energy is injected into the waves when $h$ is decreased ($\langle I \rangle \sim h$ -- see above). Indeed, experiments performed with a constant $\langle I \rangle$ forcing whatever $h$ gives similar spectra as the ones in Fig.\ \ref{fig01} and similar $f_p(h)$ evolution (see $\lozenge$-symbols in Fig.\ \ref{fig03}). This evolution is not also due to the viscous damping rate $\sim 2\nu k^2(\omega)$ \cite{Landau} or to the bottom friction one $\sim \sqrt{\nu\omega(k)}/h$ \cite{Landau} since both are decreasing functions of $h$. A possible origin of $f_p$ is the following. At the cut-off frequency $f_p$ of the capillary spectrum, the nonlinear energy transfer is changed by viscous dissipation, thus it can be assumed that the typical non linear time scale of wave interaction, $\tau_{nl}$, is of the order of the typical viscous time scale $\tau_v$: $\tau_{nl}(f_p) \sim \tau_v(f_p)$ \cite{Abdurakimov10}. Thus, if the non linear time scale depends on the fluid depth, $f_p$ estimated from this above condition should also depend on the depth. A change of the wave interaction process during the transition from a deep to a thin fluid layer may modify the nonlinear time scale that thus should depend on the depth.  However, this qualitative interpretation does not explain how the energy flux flowing across the scales accumulates at the hump scale as an analog of a bottleneck effect \cite{Ryzhenkova90}. We cannot currently measure $\tau_{nl}$ (and its possible depth dependence) with a temporal measurement of the wave amplitude at a given location, but it could be it with a spatio-temporal measurement \cite{Herbert10}. This thus would deserve further studies.

\begin{figure}[t!]
\includegraphics[width=8.5cm]{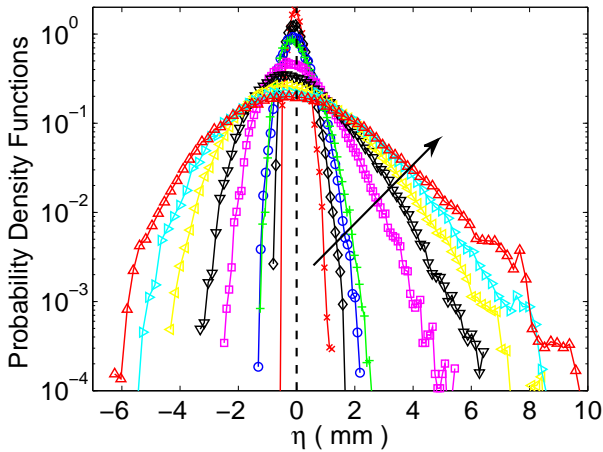}
\caption{(color online) Probability density functions of wave amplitude, $\eta$, for different values of the fluid depth, from $h =$ 3, 4, 5, 6, 8, 10, 12, 15 to 18 mm (see arrow). Forcing parameters: 1 - 6 Hz and $\sigma_V=0.8$  (arb. unit).}
\label{fig05}
\end{figure}

Fig.\ \ref{fig03} also shows the evolution of the gravity-capillary crossover frequency $f_c(h)$. For small $h$, $f_c$ is found to slightly increase with $h$, and then to saturate in the deep regime to a value known to be larger than the expected one $f_c=\frac{1}{2\pi}\sqrt{2gk_c}\simeq 17$ Hz with $k_{c} \equiv \sqrt{\rho g/\gamma}$ \cite{Falcon07}. Here, the deep limit value is $f_c\simeq 28$ Hz that corresponds to a wave number $\tilde{k}_c=1.56k_c$. By using the relation dispersion of Eq.\ (\ref{rdtheo}), $f_{c}$ then is found to be well fitted in Fig.\ \ref{fig03} by $f_c(h)=\sqrt{(g\tilde{k}_c+\frac{\gamma}{\rho}{\tilde{k}_c}^3)\tanh[\tilde{k}_c\left(h-h^*\right)]}/(2\pi)$. $h^*\simeq 2.5$~mm is the depth for which the fluid dewetting occurs on the container bottom and is of the order of the capillary length as expected for a capillary meniscus related phenomenon.  Consequently, the evolution of the crossover frequency $f_c(h)$ is given by the depth dependence of the dispersion relation at a fixed wavenumber $\tilde{k}_c$. 

\section{Distribution of wave amplitudes}
\begin{figure}[t!]
\includegraphics[width=8.5cm]{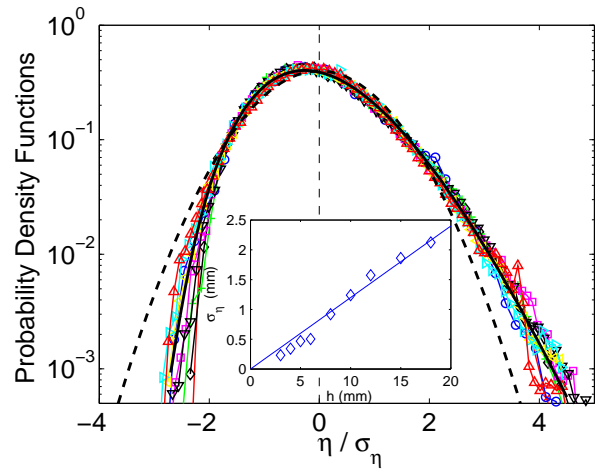} 
\caption{(color online) Same PDFs as in Fig.\ \ref{fig05} rescaled by $\sigma_{\eta}$. Dashed line: Gaussian PDF with zero mean and unit standard deviation. Solid line: Tayfun distribution with s=0.18 (see text). Inset: rms amplitude value $\sigma_{\eta}$ vs $h$. Same forcing parameters as in Fig.\ \ref{fig05}.}
\label{fig06}
\end{figure}

The probability density functions (PDFs) of the wave amplitude are shown in Fig.\ \ref{fig05} for different values of $h$ for the same value of the forcing amplitude. The PDFs are found to be asymmetrical due to strong steepness of the waves as usual in laboratory experiments \cite{Falcon07,Onorato04} or in oceanography \cite{Ochi05,Forristall00}. No dependence of this  asymmetry with the depth is observed in our experiment. Indeed, first, the skewness $S$ is roughly constant whatever the depth: $\langle S \rangle=0.55 \pm 0.1$. Second, when these PDFs are normalized to their rms amplitude values, $\sigma_{\eta}$, they all roughly collapse on a single non-Gaussian distribution as shown in Fig.\ \ref{fig06}. This means that the distribution depends only on $\sigma_{\eta}$. The bottom inset of Fig.\ \ref{fig06} shows that the rms wave amplitude $\sigma_{\eta}$ increases linearly with the depth $h$.  These rescaled distributions are found to be well fitted by a Tayfun distribution (the second-order correction to the Gaussian distribution) that reads $p[\tilde{\eta}] =\int_0^{\infty} \exp{\{[-x^2 - (1 -c)^2]/(2s^2)\}}/(\pi sc)dx$ where $c\equiv \sqrt{1 + 2s\tilde{\eta} + x^2}$, $\tilde{\eta}\equiv \eta/\sigma_{\eta}$ and $s$ the mean steepness of the waves \cite{Tayfun80}. Note that we have checked that the Tayfun distribution fit well our experimental normalized PDF performing for much larger depth (i.e., $18 \leq h \leq 140$ mm) that is for depth much larger that the wave amplitude $\eta$. This also suggests that this typical distribution does not depend on the possible change of the wave interaction process during the transition from a deep to a thin fluid layer. 

\section{Conclusion}
We have performed experiments of gravity-capillary wave turbulence on the surface of a fluid and have focused on the effects of finite depth. When the fluid depth is decreased from a deep to a thin fluid layer, a hump is observed on the wave spectrum at a depth-dependent scale located in the capillary regime. The presence of this hump strongly reduces the inertial range of the capillary wave turbulence. The mechanism of formation of this hump and its dependence on the depth are open problems. In the gravity regime, the power spectrum is still found to be a power-law but with an exponent that continuously decreases with decreasing depth in a qualitatively agreement with the weak turbulence predictions in both deep and shallow water limits. The scaling of the spectrum with the mean injected power is also measured and suggests that a change of the wave-interaction process occurs for gravity waves during the transition from a deep to a thin fluid layer. Finally, the probability distributions of the wave amplitude rescaled by its rms value are measured for different fluid depths. The distributions are found to be independent of the depth and to be well described by a Tayfun distribution whatever the fluid depth.

\acknowledgments
We thank M. Berhanu for fruitful discussions. This work has been supported by ANR Turbonde BLAN07-3-197846.


\end{document}